\documentclass{svproc}
\usepackage{url}

\usepackage{graphicx}
\usepackage{soul}

\begin{document}
\mainmatter              
\title{Comparative study of variational quantum circuit and quantum backpropagation multilayer perceptron for COVID-19 outbreak predictions}

\author{Pranav Kairon\inst{1} \and Siddhartha Bhattacharyya\inst{2}}
\institute{Delhi Technological University, Bawana, Delhi, India,\\
\email{pranavkairon\_bt2k17@dtu.ac.in}
\and
CHRIST (Deemed to be University), Bangalore, India, \\
\email{dr.siddhartha.bhattacharyya@gmail.com}}

\maketitle              

\begin{abstract}
There are numerous models of quantum neural networks that have been applied to variegated problems such as image classification, pattern recognition etc.Quantum inspired algorithms have been relevant for quite awhile. More recently, in the NISQ era, hybrid quantum classical models have shown promising results. Multi-feature regression is common problem in classical machine learning. Hence we present a comparative analysis of continuous variable quantum neural networks (Variational circuits) and quantum backpropagating multilayer perceptron (QBMLP). We have chosen the contemporary problem of predicting rise in COVID-19 cases in India and USA. We provide a statistical comparison between two models , both of which perform better than the classical artificial neural networks.
\end{abstract}

\section{{\large{}{}Introduction}}

The Novel Corona virus disease/COVID-19 was first detected in Wuhan, China on 31st December 2019. After the 1918 H1N1 influenza, COVID-19 has been accounted for as the most pernicious respiratory infection to affect the population worldwide \cite{Darwish2020}. With meteoric spread it has taken lives of half a million people and around 13 million confirmed cases bringing lives of millions to a standstill. According to the World Health Organization report 170 countries now have adumbrated at least one case as on July, 2020. While, 4.5\% is the mortality rate of this deadly disease, for the age group 70-79 this has ascended to be 8.0\% while for those above 80 it is soaring 14.8\%. People already suffering from heart disease, diabetes are especially at higher risk, the fact that they are older than 50 only exacerbates the situation  \cite{Zheng2020}. Middle East Respiratory Syndrome (MERS), Severe Acute Respiratory Syndrome (SARS) and COVID-19 all belong to a family of severe acute respiratory syndrome, called as coronavirus. The symptoms are self evident within 2-14 days \cite{Bai2020} and have  been signified by gamut of fever, cough, shortness of breath to pneumonia. Flattening the epidemic curve is key part in managing this pandemic since there is no commercially available vaccine against COVID-19. Although there is some hope as researchers in Russia and Oxford University have completed the human trials of the vaccine, which show positive results. \\
Given the effectual predictions in healthcare, machine learning has proven itself fundamentally. After intensifying patterns and expertise of radiologists, machine learning is now helping them to predict the disease and diagnose it earlier. Also with availability of meager data on COVID-19 the role of data scientists has increased drastically  Since they have to integrate the data and then release the results which will help in taking unerring decisions.~\cite{Koike2018}.\\
Quantum computing \cite{Steane1998} is an amalgamation of quantum physics and computer science. Originally proposed in the 1970s, the research remained mostly theoretical until early 2000s with advent of Shor's and Grover's Algorithms \cite{Monz2015}. We are presently in the NISQ (Noisy intermediate Quantum) era of quantum computing which implies that the presently available quantum processors are small and noisy. Since many countries are investing heavily on quantum technologies, the growth rate has been accelerated in past two years. Quantum computers supersede classical computers by harnessing quantum effects for computational vantage providing polynomial and even exponential speedups for specific problems. With Moore's law coming to an end, quantum computers become even more eminent since they are not made up of small transistors but rather quits. On the other hand we are living in an age of data and machine learning provides robust models for predictions, classification and organisation tasks.

\section{Motivation}

Presently available artificial neural network models provide robust models for plethora of problems ranging from stock prediction and statistical modeling to image recognition, data mining \cite{Huarng2006}. However, quantum neural networks are still somewhat unexplored in terms of their applicability to eclectic and pragmatic problems. In terms of convergence rate and fit ability, Quantum Backpropagation Neural Networks have outperformed classical ANNs, since they can harness the mathematical advantage of complex numbers \cite{Mitrpanont2004} using Quantum phase
door and two negates controlled controlled Quantum neuron model. Continuous variable Quantum neural networks \cite{Killoran2019} are similar to variational circuits in that they are free parameter dependent quantum algorithms. They are gaining attention as an alternate way to characterize NISQ era devices. Given the guessing abilities of quantum neural systems, these networks are agile to cost degradation.These properties have roused us to apply these network models to regressional tasks like predicting rise in cases for COVID 19 and compare them against the intelligent paradigmatic classical ANNs.

\section{Proposed Methodology}

The experimental data has been collected from \mbox{\cite{hde2019}} for India and USA. We have considered a live time-series dataset from $30^{th}$ January till $1^{st}$ July 2020. It provides the number of confirmed cases, number of deaths and number of people who have recovered from the pervasive infection. We can find additional two features i.e. the number of active cases which is given by Active cases=No.of Confirmed Cases - (No.of deaths + No. of recovered patients) and New cases added each day which is just the difference of active cases on the $n$$^{th}$ day and $(n-1)$ $^{th}$ day. The global data for all the three columns is plotted in Figure 1. The pervasive nature of the infection is evident from the exponential curves in Figure 1. In this work, Quantum Neural Networks (QNNs) have been employed to predict rise in cases. QNNs are seldom applied to regressional tasks \cite{Diep2020}. We consider two kinds of quantum neural networks. First being, a Quantum Backpropagation Multilayer Perceptron (QBMLP) which utilizes the superposition feature and complex numbers~\cite{Mitrpanont2004}\cite{Bhattacharyya2015}. Secondly, we employ a Continuous Variable Quantum Neural Network (CVQNN) provided by Xanadu in their Pennylane and Strawberry fields package \cite{Killoran2019}. CVQNNs are variational circuits that use Gaussian and non-Gaussian gates to perform affine and non linear transformations. Both the QBMLP and CVQNN can be generalized to any number of hidden layers, and are characterized by the Tanh and Kerr activation functions, respectively. The accuracy for both the models in finding out the rise in number of cases and deaths due to COVID-19 in India and USA is tested. Moreover, a comparative study has been made between both the networks using a two tailed statistical $t$-test. CVQNN outperforms QBMLP in some cases whereas an opposite effect is seen in some, as can be seen from the two tailed $t$-test. 
\begin{figure}
	\begin{tabular}{cc}
		\includegraphics[scale=0.25]{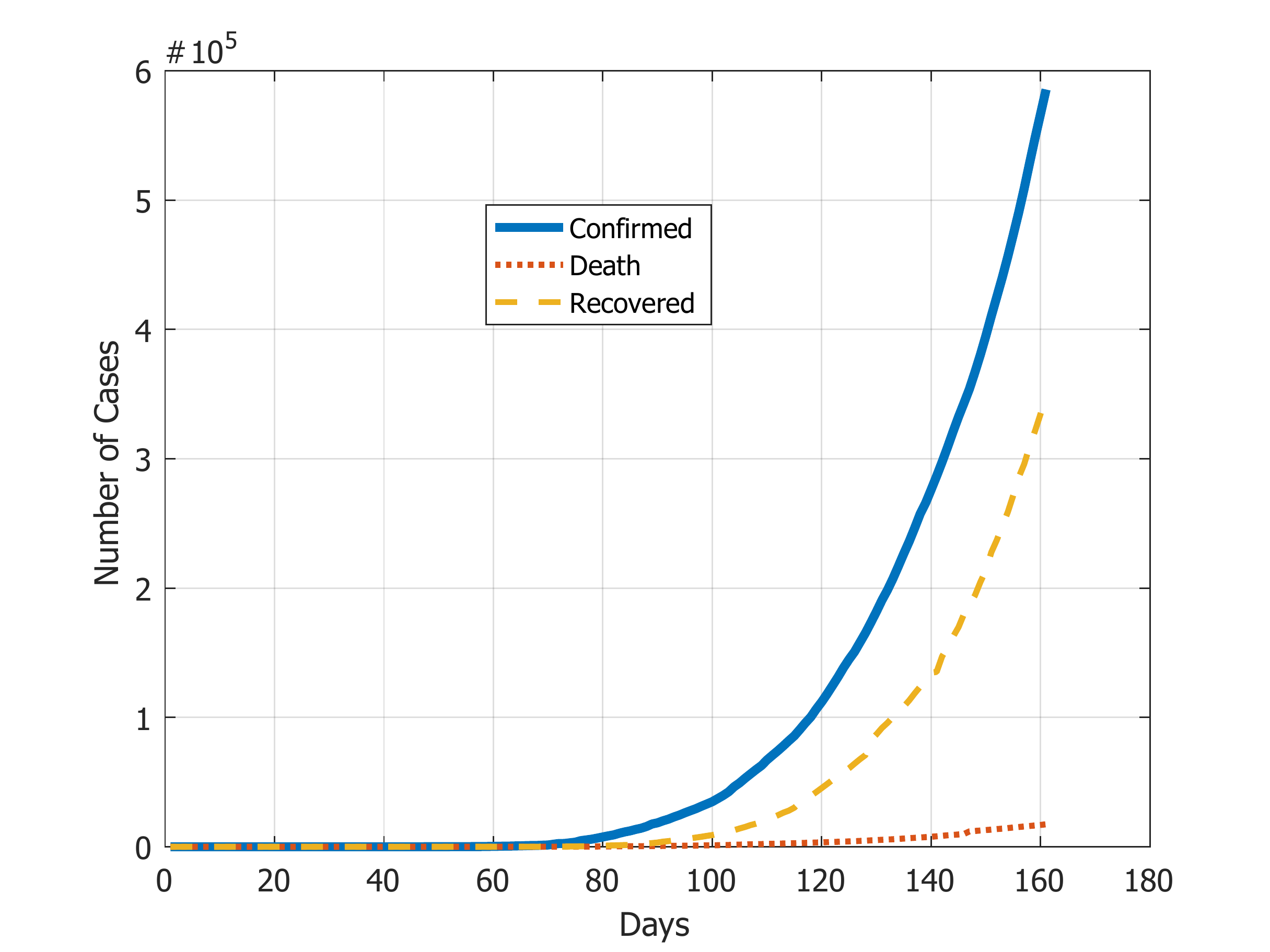} & \includegraphics[scale=0.25]{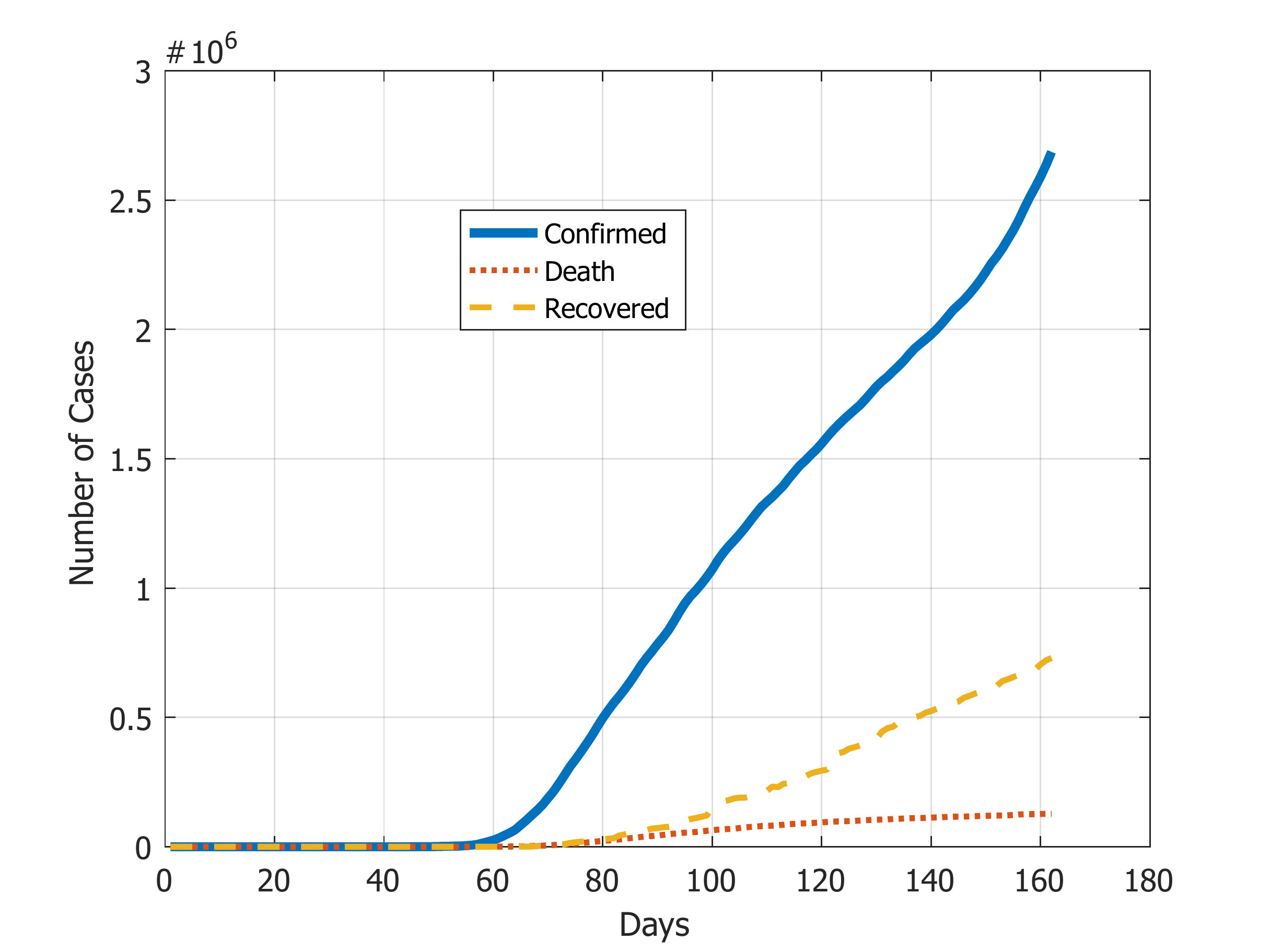}\tabularnewline
		(a) India COVID-19 cases & (b) USA COVID-19 cases\tabularnewline
	\end{tabular}
	
	\caption{\label{fig:1}Total number of cases for all the countries from January
		$30^{th}$ to $1^{st}$ July 2020}
	
\end{figure}

\section{Models} 

Deep feed forward neural networks have shown tremendous results on problems like image processing, stock predictions etc. They owe their success to affine transformations done through hidden layers, which are the building blocks of a mulitlayer perceptron, and activation functions that introduce non linearity to map input $x$ to output $y$. A typical architecture can be represented by
\begin{equation}
L:x\rightarrow y=\phi(Wx+b)
\end{equation}
for $l$ input variables and $m$ output variables, $W$ is a $l\times m$ matrix and $b$ is constant vector of length $l$, where $W, b~\epsilon R$. $\phi$ is usually chosen from a pool of functions, it can be sigmoid, tanh, $ReLU(x)=max(0,x)$. Inputs for all the layers form the output for the next layers, and the weights are optimized for the defined task on basis of hyper parameters adjusted by the user, which include the number of layer wise input variables (features) and number of layers ($depth$ $d$).

\subsection{Fuzzified Quantum Backpropagation Multilayer Perceptron}

Consider a dataset of $N$ features that can be represented as $x=(x_{1,}x_{2,}....x_{N})$. The input data can be squashed into the range $[0,1]$ using fuzzy linear membership function \cite{Bhattacharyya2015} with 1 signifying the highest value whereas $0$ lies on opposite end of the spectrum. Membership value is allotted to all the data points in between using linear scaling rule~\cite{Mitrpanont2004}. \\
Given the inputs $x$, we define a function $f(x)$ as 
\begin{equation}
\left|x\right\rangle =cos(x)\left|0\right\rangle +sin(x)\left|1\right\rangle \equiv cos(x)+i.sin(x) \label{eq:2}
\end{equation}
The weights which are modeled as rotation gates, $R(\theta_{i})$ to the input qubits, $\left|x_{i}\right\rangle $ with input bias $\lambda$. Hence, the modified neural network equation takes the form 
\begin{equation}
u=\sum_{i=1}^{n}R(\theta_{i})\left|x\right\rangle -\left|\lambda\right\rangle \label{eq:3}
\end{equation}
Now, instead of applying just a non linear function as in the classical case, the transformed output is given by
\begin{equation}
y=\frac{\pi}{2}g(\delta)-\arctan(\frac{amplitude(\left|1\right\rangle )}{amplitude(\left|0\right\rangle )})\label{eq:4}
\end{equation}
where, $g(\delta$)=$\frac{1-e^{-2\delta}}{1+e^{-2\delta}}$ is the tanh function. \\
Similar to the classical Multilayer Perceptron, the QBMLP is designed with the first layer accepting inputs in form of qubits, followed by any number of hidden layers, ending at an output layer with a single neuron (for regressional tasks). The qubits for QBMLP are designed by transforming the squashed input $x$ into the range of $[0,\frac{\pi}{2}]$ so that they become suitable angles for rotation gates, which after can be transformed to
\begin{equation}
\left|\frac{\pi}{2}x_{i}\right\rangle =cos(\frac{\pi}{2}x)\left|0\right\rangle +sin(\frac{\pi}{2}x)\left|1\right\rangle \label{eq:5}
\end{equation}
The output of the model is a quantum state expressed as a complex number in Euler's form. To extract the physical meaning of this output, either a quantum measurement can be done and the probability of getting $\left|0\right\rangle,\left|1\right\rangle$ is considered as the output or equivalently the final output can be obtained as square of either the real or the imaginary part of the output. In this paper, we have taken complex part of output which is the same as taking probability of getting $\left|1\right\rangle$ as the final result. The necessary changes need to be made in the analysis of backpropagaton if one decides otherwise.
\begin{equation}
\mathcal{O}=Probability(\left|1\right\rangle )=\mid Im(y)\mid^{2}=sin^{2}(y)\label{eq:6}
\end{equation}
Interested readers may refer to~\cite{Bhattacharyya2015} for details on the architecture and operation of the network architecture.

\subsection{Continuos Variable Quantum Neural Network}

Pennylane is a quantum macine learning python library provided by Xanadu\cite{Bergholm2018}, working on the principle of quantum differentiable
programming. It provides an efficient platform by combining quantum simulators and classical machine learning, that helps users in traning various circuits. \cite{Killoran2019a}. The Strawberry fields in a photonic quantum computing library used to solve plethora of problems such as boson sampling, graph optimiation etc. This has been used to construct a photonic neural network model having continuos variable gates. Variational circuits \cite{Liu2020} behave similar to neural networks in that, there is a definite input (input quantum state) which is embedded into the circuit using a suitable embedding, weights (circuit parameters) that need to be learnt by the model and quantum measurement of an observable say $A$, which generates a classical output on which training rules are applied for parametric learning. A three qumode CVQNN architecture is shown in Figure 2 (a). They are quite useful algorithms for NISQ devices in prblems such as quantum chemistry, optimization etc. \\
This continuous variable quantum neural network actually contains classical neural networks within itself. The use of continuous variable Gaussian and non Gaussian operations such as displacement operators, squeezing gates, Kerr gates etc, attribute novelty to this approach given the fact that it doesn't contain quantum features such as entanglement and superposition in contrast with other variational quantum circuits. If the input dataset has $N$ features (columns),
it can be encoded into the circuit using displacement operators. Let
the classical input $x$ is given by $x=(x_{1},x_{2}...x_{N})$
and the initial quantum state of circuit is given by ${\left|x\right\rangle =\bigotimes_{i=1}^{N}}{\left|x_{i}=0\right\rangle }$. Then the displacement operators, $D(x)$ can be applied to all qumodes, to encode classical input $x$ such that 
\begin{equation}
|x\rangle =\bigotimes_{i=1}^{N}D(x_{i})|x_{i}=0\rangle
\end{equation}
Just like in classical feed forward neural networks, any layer takes inputs from the output of the preceding layer, results are extracted by applying homodyne measurements on each qumode. A similar type of transformation has been used, by using continuous variable quantum analogs. Information about the various CV analogs applied in the quantum neural network can be found in Table 1. Combining all of the above, in order to obtain the our full affine transformation $D\circ U\circ S\circ U\left|x\right\rangle =\left|W{x+d}\right\rangle$, we can simulate a classical neural network layer which if broken down is basically a transformation given by equation \ref{eq:12}, and the post proccessing of the output is similar to what is done in all neural network models, i.e. paramters are updated on basis of training rules as
\begin{equation}
L(z)=\phi(Mz+\alpha)
\label{eq:12}
\end{equation}

\begin{table}
	\begin{centering}
			\caption{Various CV quantum gates used to simulate affine transformation}
		\begin{tabular}{ccc}
			\hline 
			Classical  & CV analog & Transformation \tabularnewline
			\hline 
			Weight Matrix : W & Interferometer :$\hat{\mathcal{U}}$ & $\hat{\mathcal{U}}|x\rangle =\bigotimes_{i=1}^{N}|\sum_{j=1}^{N}C_{ij}x_{j}\rangle =|C{\bf x}\rangle$\tabularnewline
			& Squeezing gate : $\hat{\mathcal{S}}$ & $\hat{\mathcal{S}}(r)\left|x\right\rangle =e^{-\frac{1}{2}\sum_{i}r^{i}\left|\sum x\right\rangle }$\tabularnewline
			Bias : b & Displacement Operator :$\hat{\mathcal{D}}$ & $\hat{\mathcal{D}}(\alpha)\left|x\right\rangle =\left|x+\alpha\right\rangle $\tabularnewline 
			Activation Function : $\phi$ & Non-Gaussian transformation : $\Phi$ & $\phi\ensuremath{|x\rangle}=|\Phi(x)\rangle$\tabularnewline
			\hline 
		\end{tabular}
		\par\end{centering}

\end{table}

\begin{figure}
	
	\begin{tabular}{|c|c|}
		\hline 
		\includegraphics[scale=0.25]{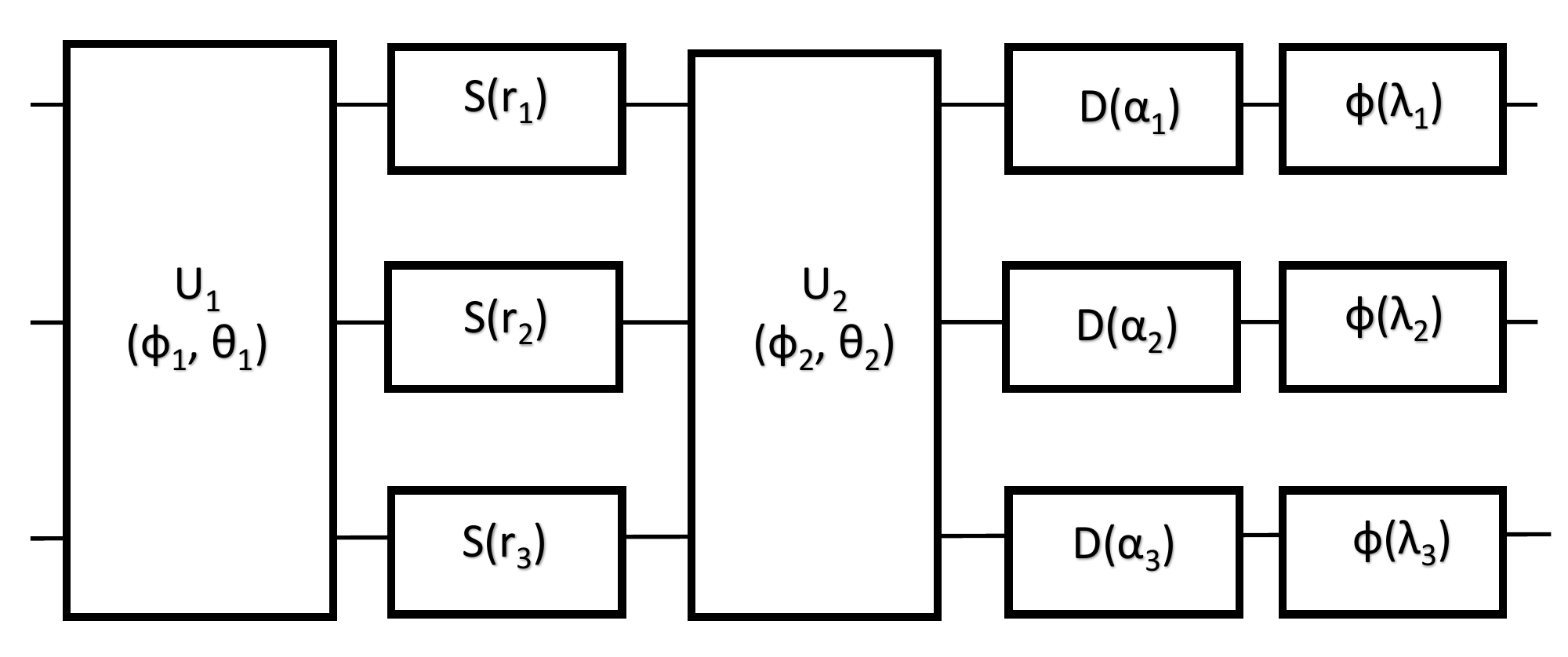} & \includegraphics[scale=0.25]{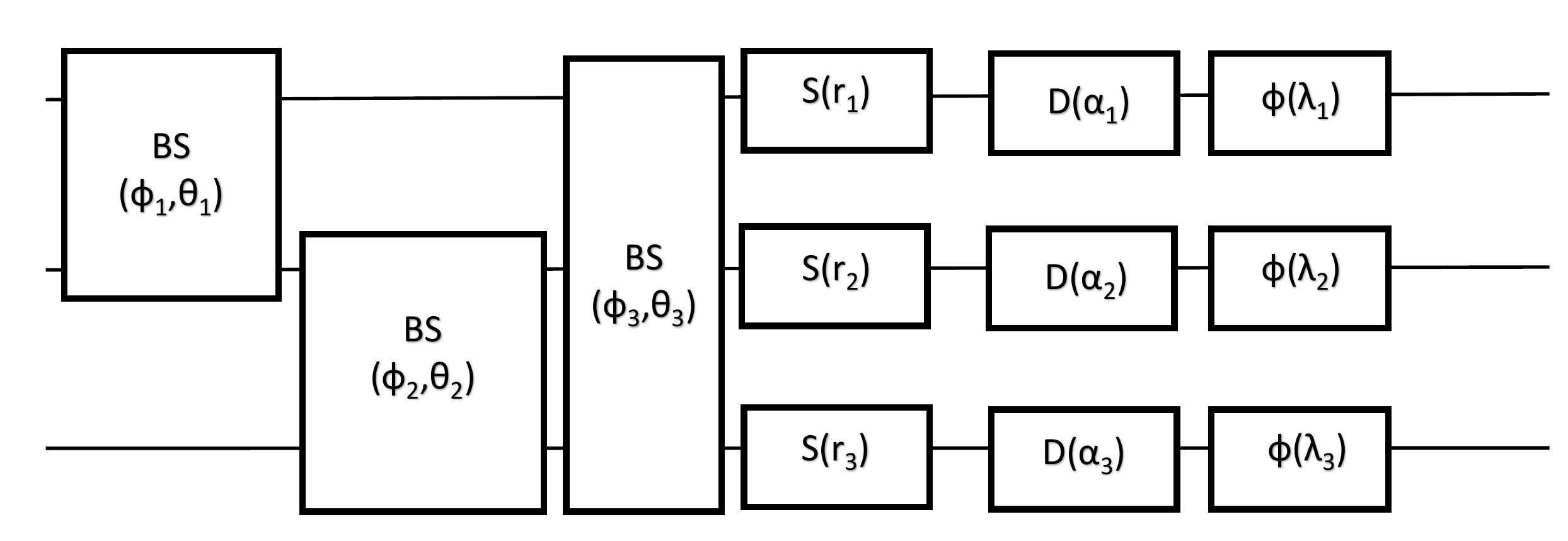}\tabularnewline
		\hline 
		(a) Standard CVQNN architecture  & (b) Modified CVQNN architecture\tabularnewline
		\hline 
	\end{tabular}
	
	\caption{Three qumode CVQNN architectures}
	
\end{figure}

\section{Experimental Results} 

Both the QBMLP and CVQNN have been applied to study the rise of corona virus cases in India and USA. The problem is same as multi dimensional regression analysis done commonly using classical neural nets. The networks have been trained to map non linear functions on basis of given data points and their performance has been compared on basis of two tailed $t$-test. A 3-3-1 architecture has been designed for QBP (i.e. 3 input neurons, 3 hidden neurons and 1 output neuron) and for the CVQNN, a 3 qumode quantum circuit has been constructed with only one hidden layer shown in Figure 2 (a). The output of the first qumode is finally measured using homodyne measurement in position eigenstates. The four parameters used from the data are Confirmed cases, Number of Deaths and Recoveries and number of days passed. Three of these features have been used to predict confirmed cases and the number of deaths since this information is quite important with regards to public safety and social distancing standards. The activation function used is $tanh$ as opposed to the more commonly used sigmoid function for quantum inspired model. The learning rate has been defined to be 0.001 for 15000 iterations($I_{QBMLP}$). Since a nested for loop has been used in the algorithm of QBMLP, it's time complexity is given by $\mathcal{O}(Len(y)*I_{QBMLP})$, where Len(y) is specified by the length of output column. For CVQNN, the learning rate has been chosen as 0.1 (using stochastic gradient descent) with 200 iterations($I_{CVQNN}$). The Strawberry fields simulator which is used to perform the CV computation, utilizes time/memory as $\mathcal{O}(C^{W}*I_{CVQNN})$ where $C$ is the cutoff dimension (a hyper parameter, equal to 10 for our case) and $W$ is the number of wires (or qumodes) involved in the computation which were 3 in this case, as their were 3 features. The caveat is, if we choose cutoff dimension to be low in order to speed up the computation, some operations like Displacement, Squeezing, Kerr operations push the quantum states out of the defined space giving low fidelity. Hence, the standard layer architecture has been modified to a novel version to speed up the computation as shown in Figure 2 (b). As we can see from the algorithm the CVQNN has better time complexity than our QBMLP model, but owing to sluggish nature of Strawberry fields simulator the former takes longer to execute in real time than the latter. The prediction results and the cost function decay for QBMLP and CVQNN for the test data set are shown in Figures 3 and 4, respectively.

\begin{table}
	\begin{centering}
		\caption{Two tailed $t$-test for QBMLP and CVQNN}
		\begin{tabular}{|c|c|c|c|c|}
			\hline 
			Data Set/Model & \multicolumn{1}{c}{QBMLP} &  & \multicolumn{1}{c}{CVQNN} & \tabularnewline
			\cline{2-5} 
			& Statistic & $p$ value & Statistic & $p$ value\tabularnewline
			\hline 
			India - Death Prediction & 0.0572 & 0.9545 & 0.01998 & 0.98411\tabularnewline
			\hline 
			India - Confirmed Prediction & -0.0189 & 0.9849 & 0.06372 & 0.94939\tabularnewline
			\hline 
			USA - Death Prediction & -0.02261 & 0.9820 & -0.01964 & 0.98438\tabularnewline
			\hline 
			USA - Confirmed Prediction & 0.0612 & 0.95132 & 0.08899 & 0.92936\tabularnewline
			\hline 
		\end{tabular}
	\end{centering}
\end{table}

\section{Discussions and Conclusion}

In this study the ability of quantum neural networks to perform regressional tasks has been investigated, specifically in predicting the rise in COVID-19 cases. The operational features used are the number of days passed, number of confirmed, recovered and death cases. The performance of a Quantum Backpropagation Multilayer Perceptron and a Continuos Variable Quantum Neural Network has been compared for the above mentioned task on the basis of a two tailed statistical $t$-test and it is found to be almost similar. While the QBMLP model performs better for checking the rise of confirmed cases in both the countries, the CVQNN model outperforms the former for calculating the rise in the number of deaths. Although the difference in $p$ values for both the models is not that significant, it provides an interesting insight into the nature of both the models.

\begin{figure}
	\begin{centering}
		\begin{tabular}{cc}
			\includegraphics[scale=0.25]{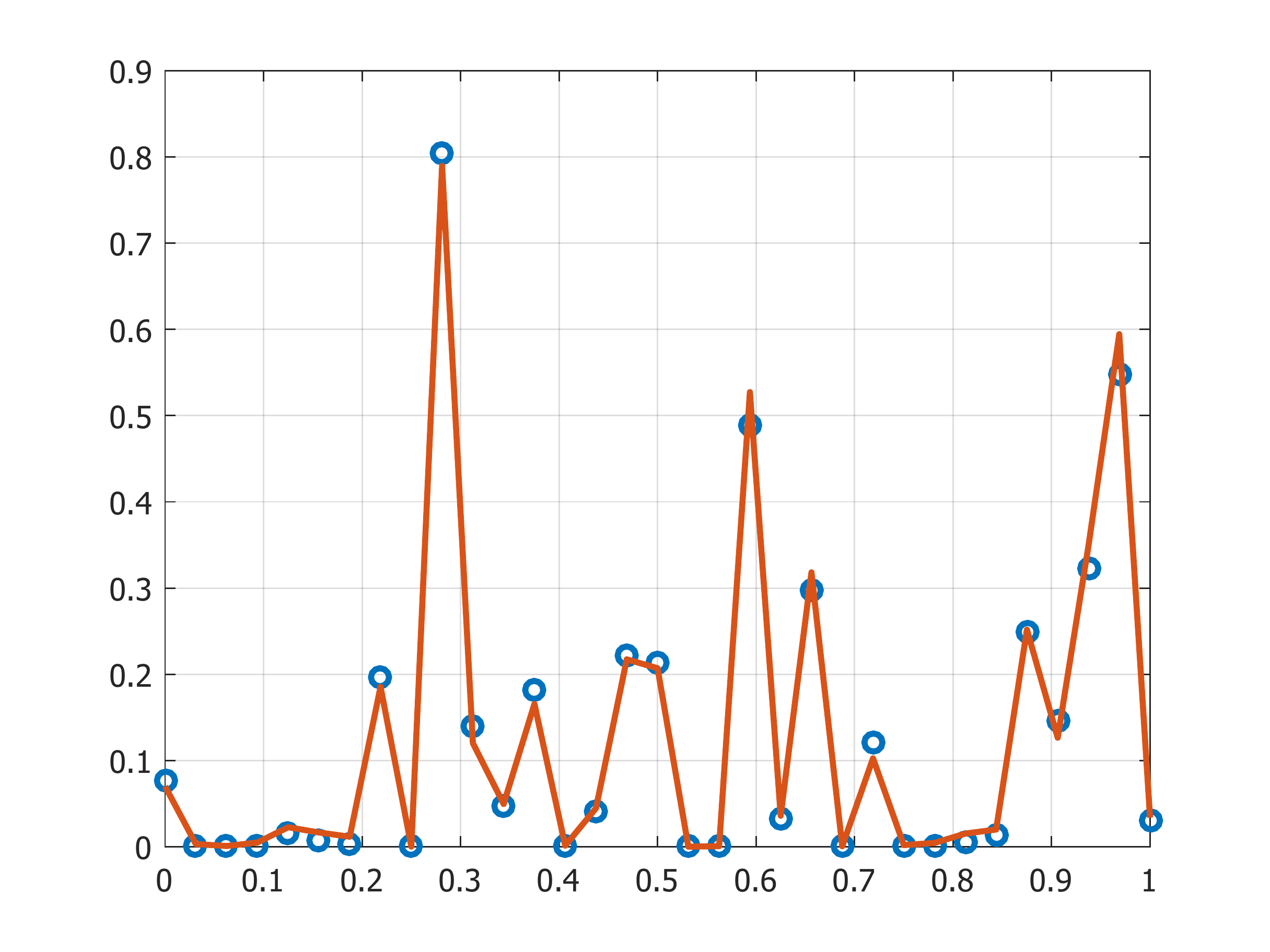} & \includegraphics[scale=0.25]{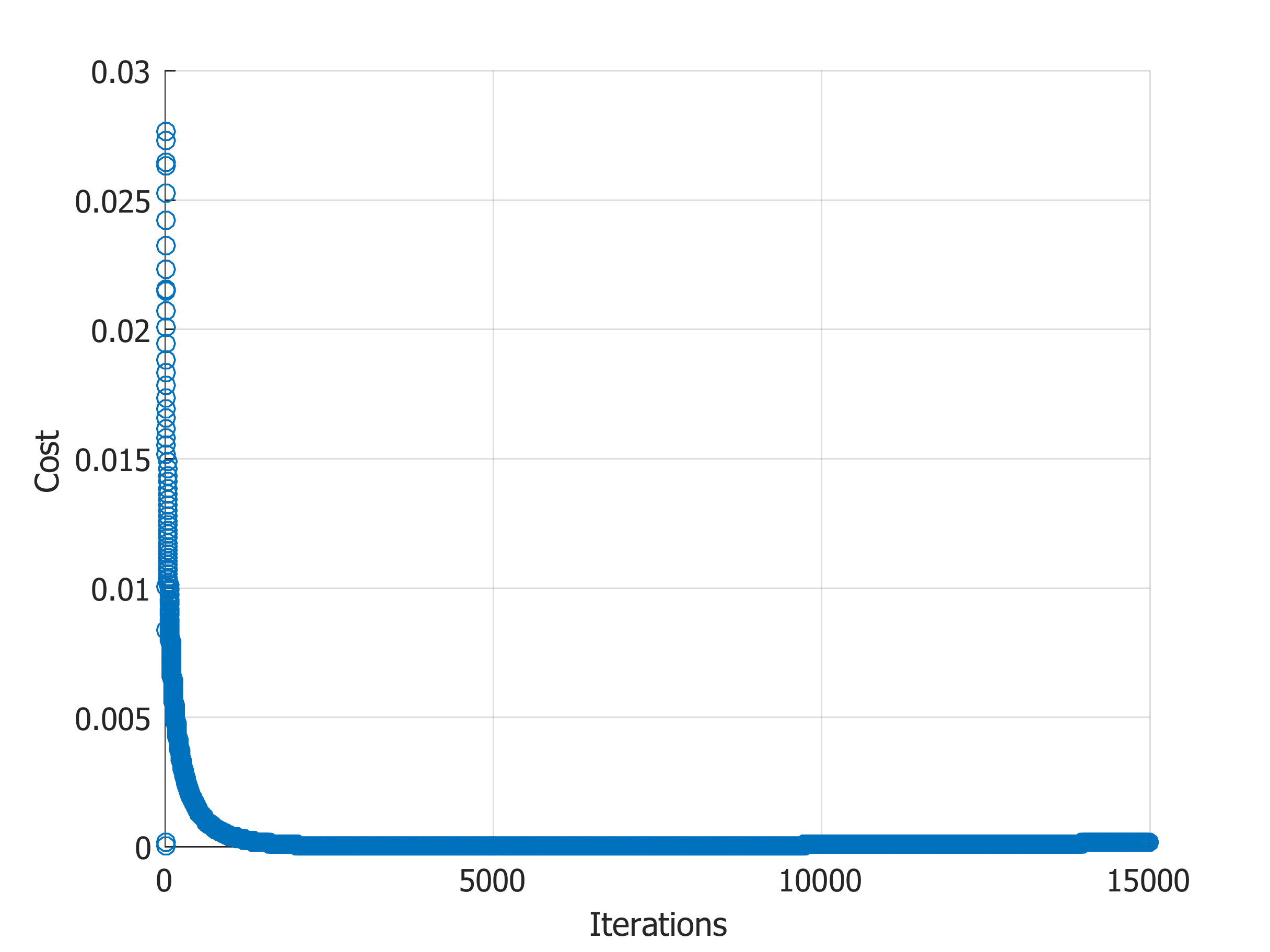}\tabularnewline
			(a) India : Death cases & (e) Cost function decay (a)\tabularnewline
			\includegraphics[scale=0.25]{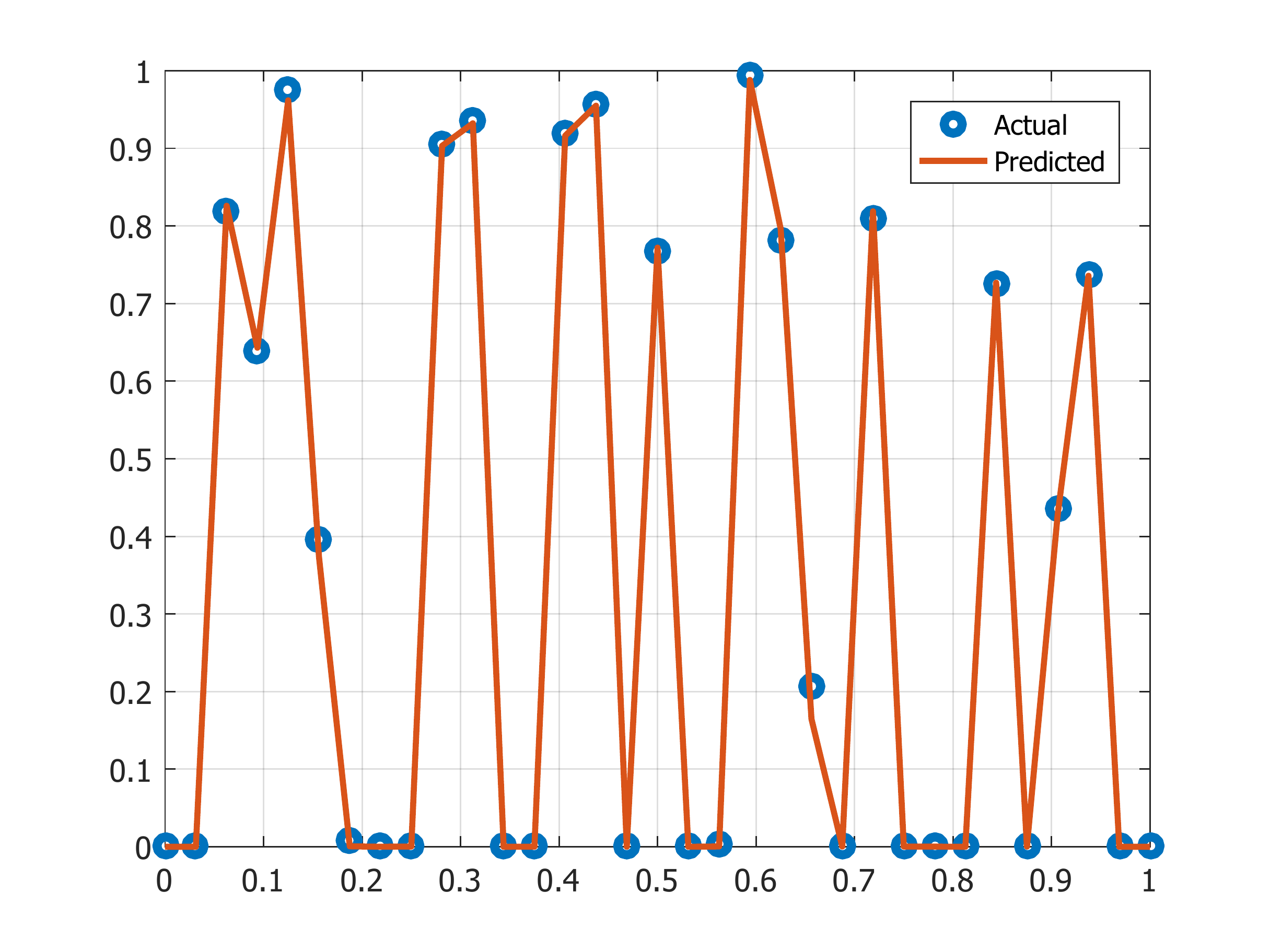} & \includegraphics[scale=0.25]{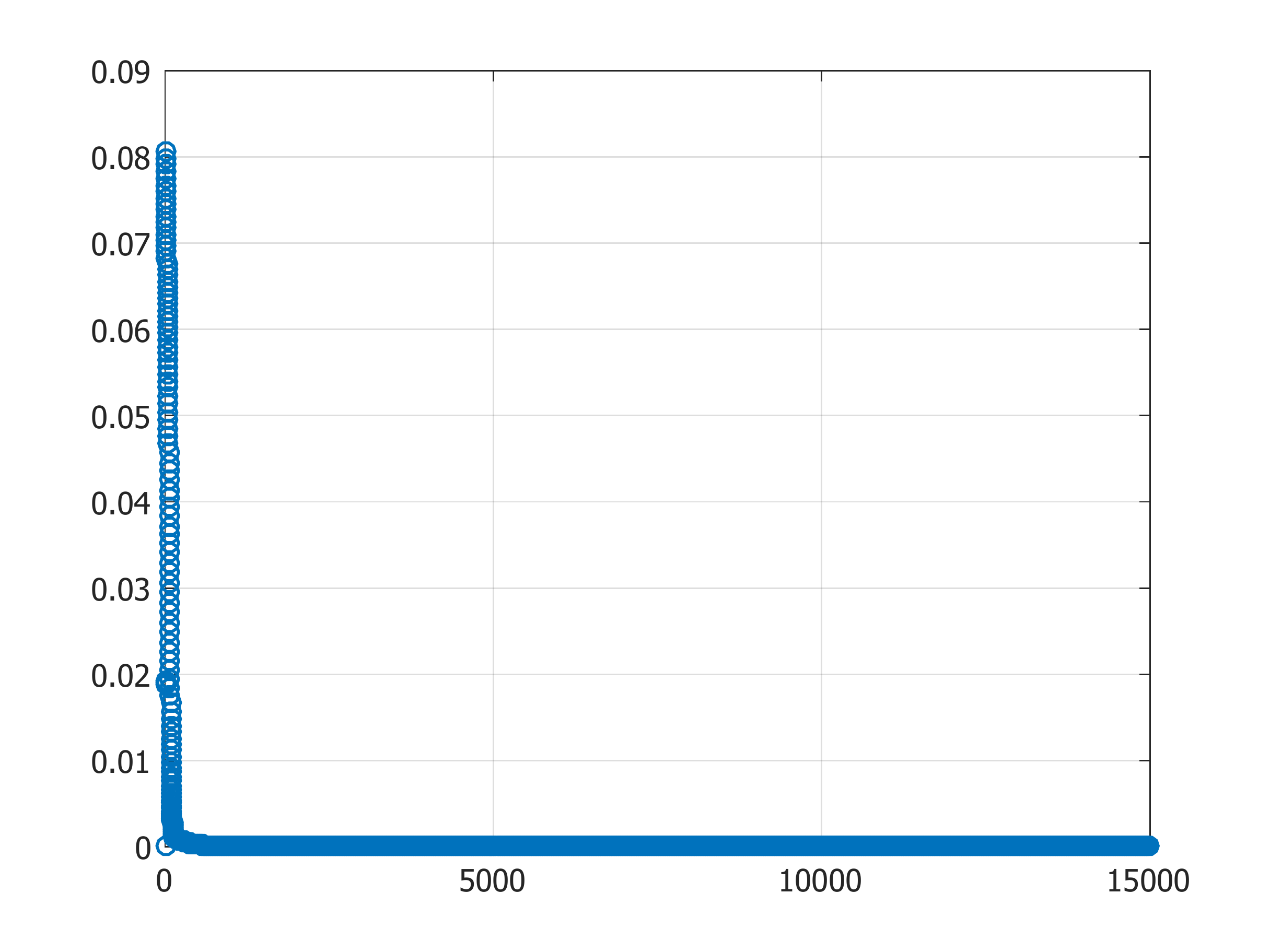}\tabularnewline
			(b) USA : Death cases & (f) Cost function decay (b)\tabularnewline
			\includegraphics[scale=0.25]{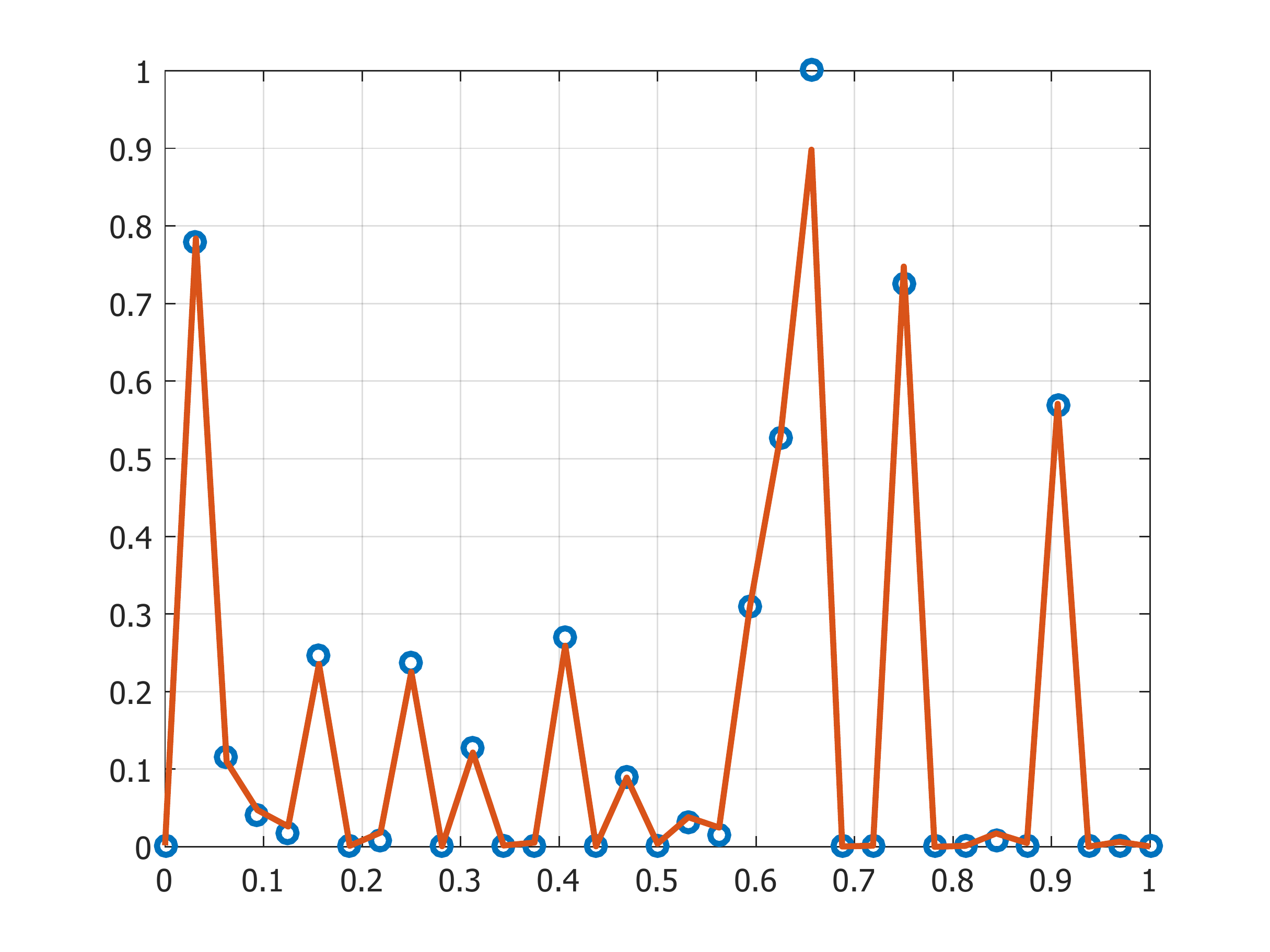} & \includegraphics[scale=0.25]{images/indeQcost}\tabularnewline
			(c) India : Confirmed cases & (g) Cost function decay (c)\tabularnewline
			\includegraphics[scale=0.25]{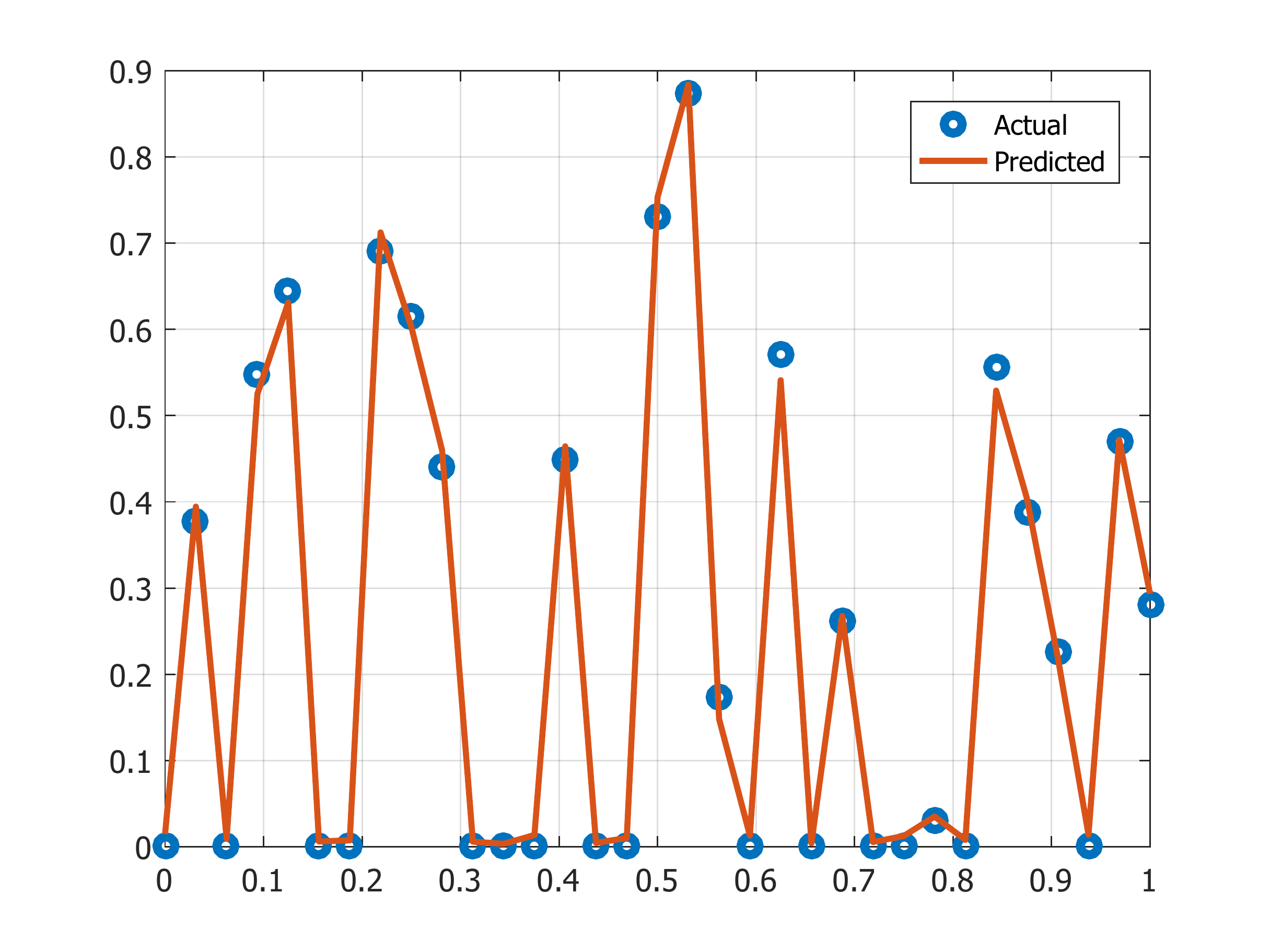} & \includegraphics[scale=0.25]{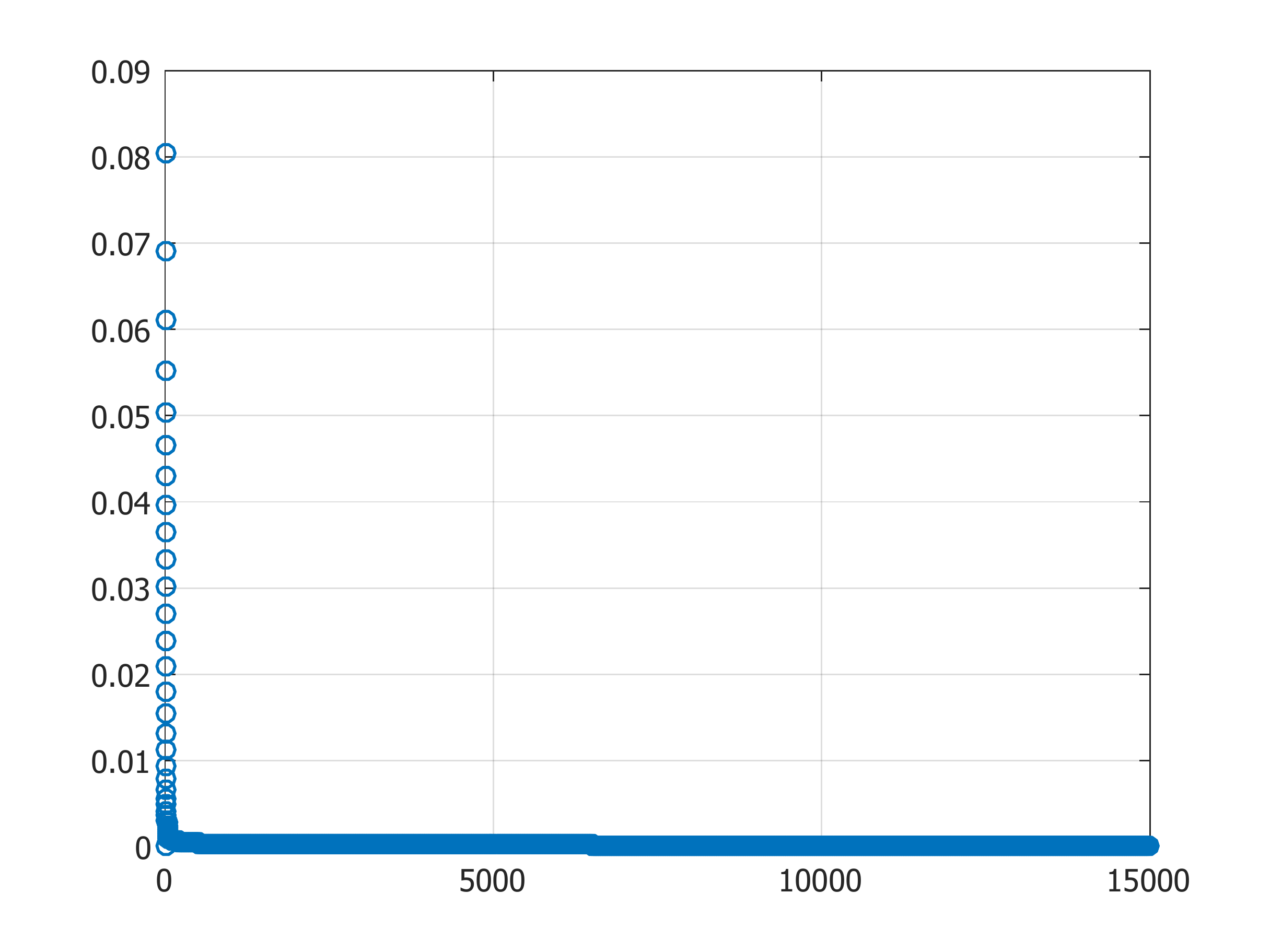}\tabularnewline
			(d) USA : Confirmed cases & (h) Cost function decay (d)\tabularnewline
		\end{tabular}
		\end{centering}
	\caption{Results and cost function decays of QBMLP for number of deaths and
		confirmed cases for COVID-19}
	
\end{figure}

\begin{figure}
	\begin{centering}
		\begin{tabular}{cc}
			\includegraphics[scale=0.25]{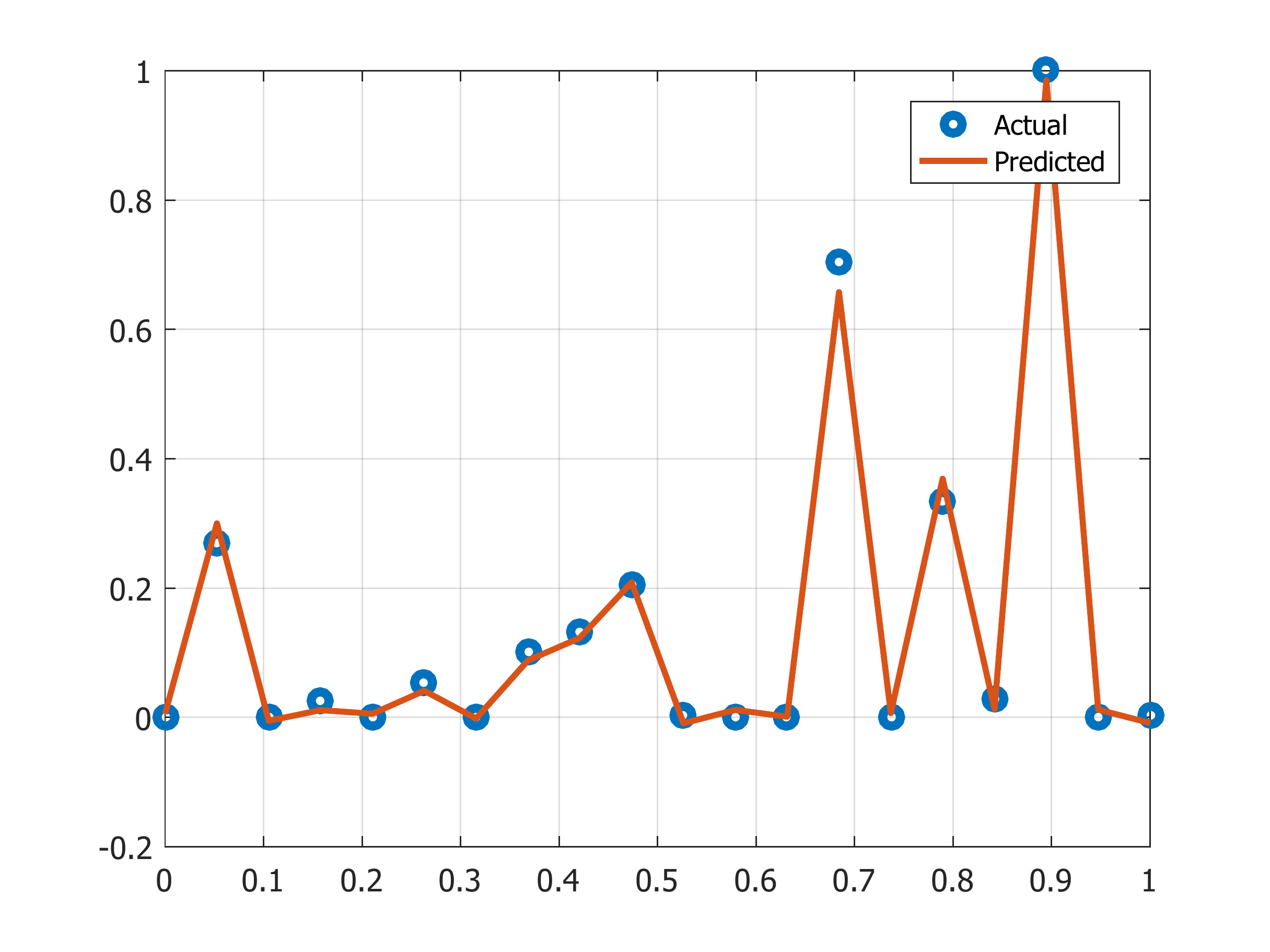} & \includegraphics[scale=0.25]{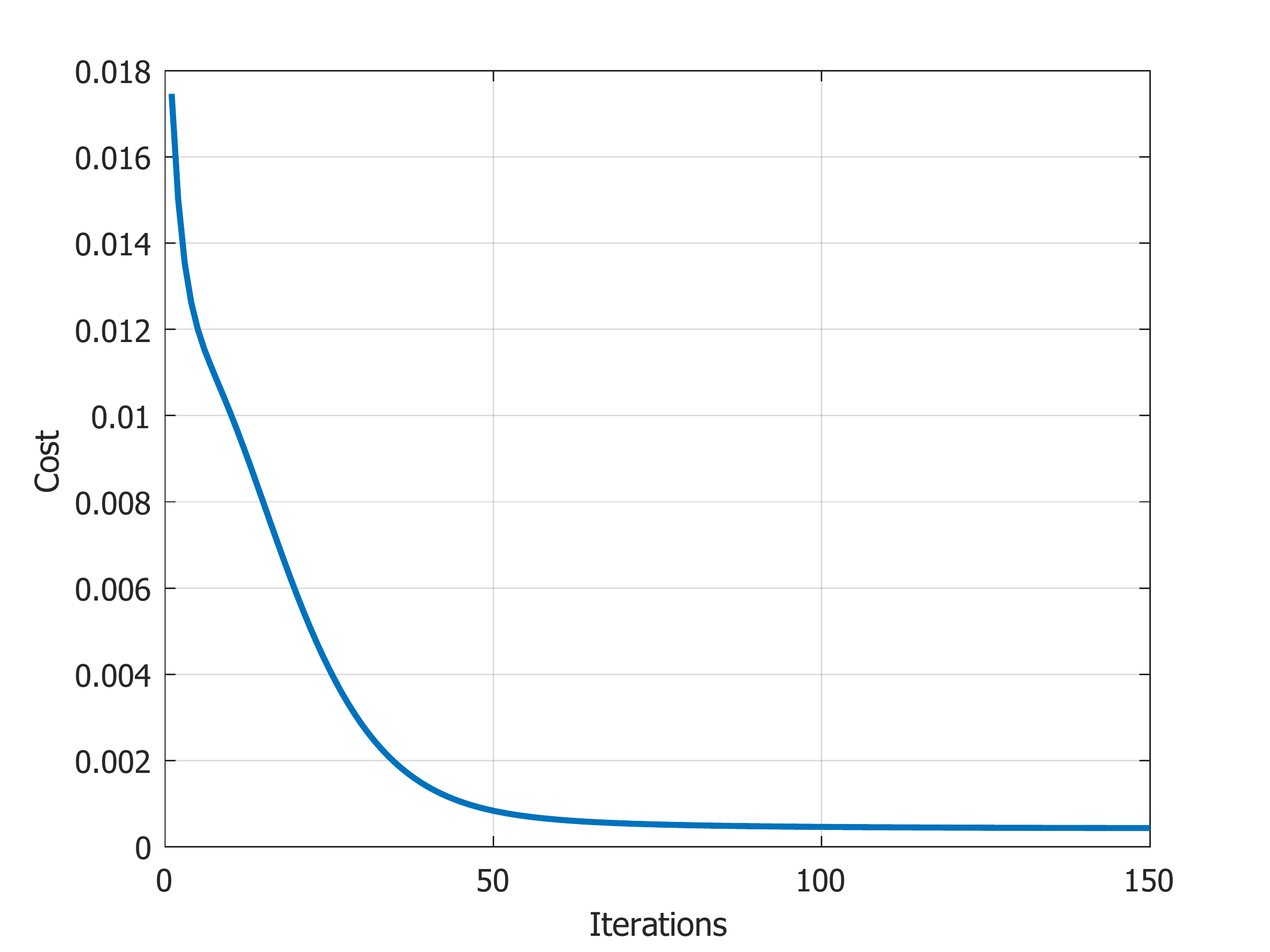}\tabularnewline
			(a) India : Death cases & (e) Cost function decay (a)\tabularnewline
			\includegraphics[scale=0.25]{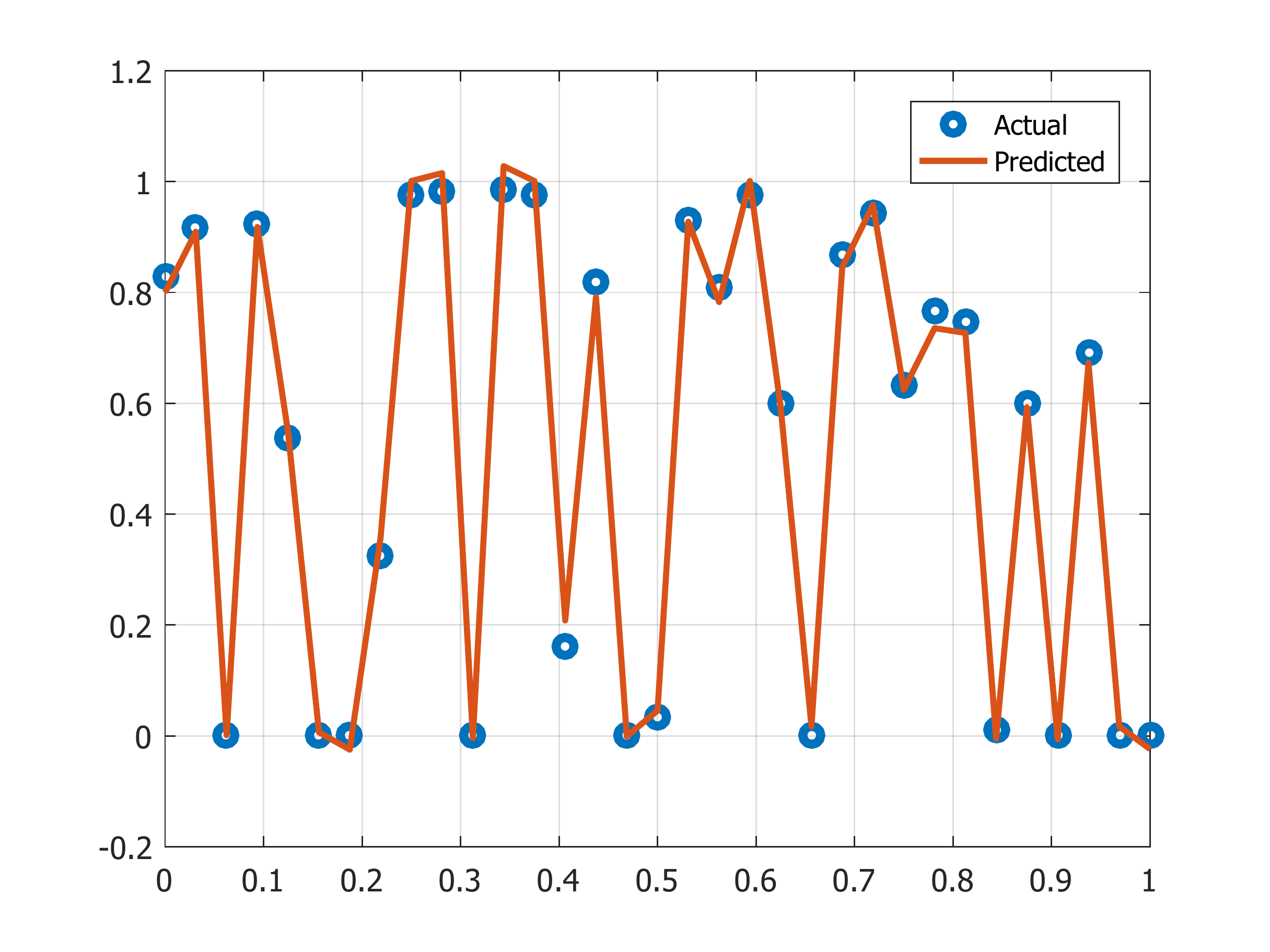} & \includegraphics[scale=0.25]{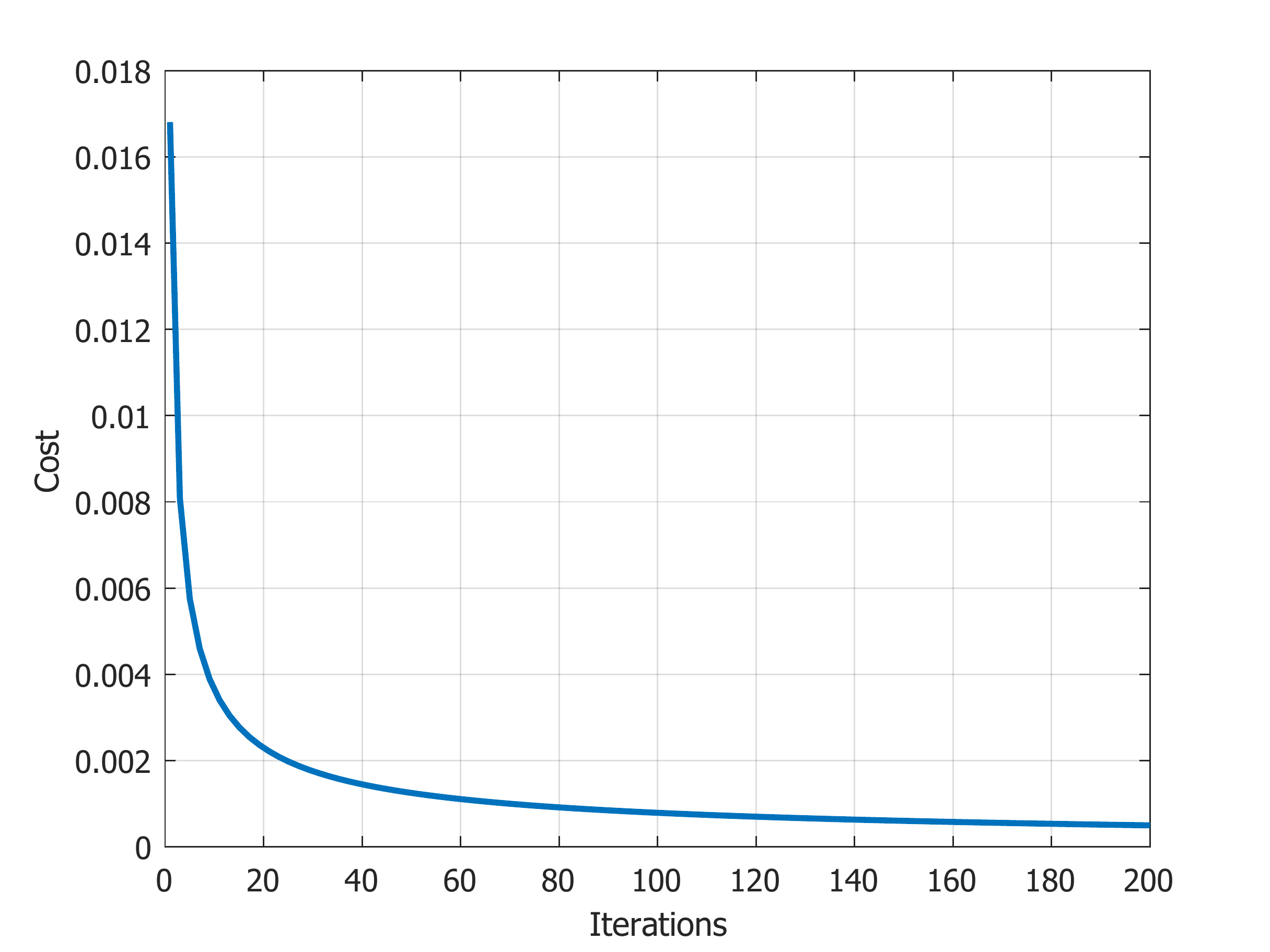}\tabularnewline
			(b) USA : Death cases & (f) Cost function decay (b)\tabularnewline
			\includegraphics[scale=0.25]{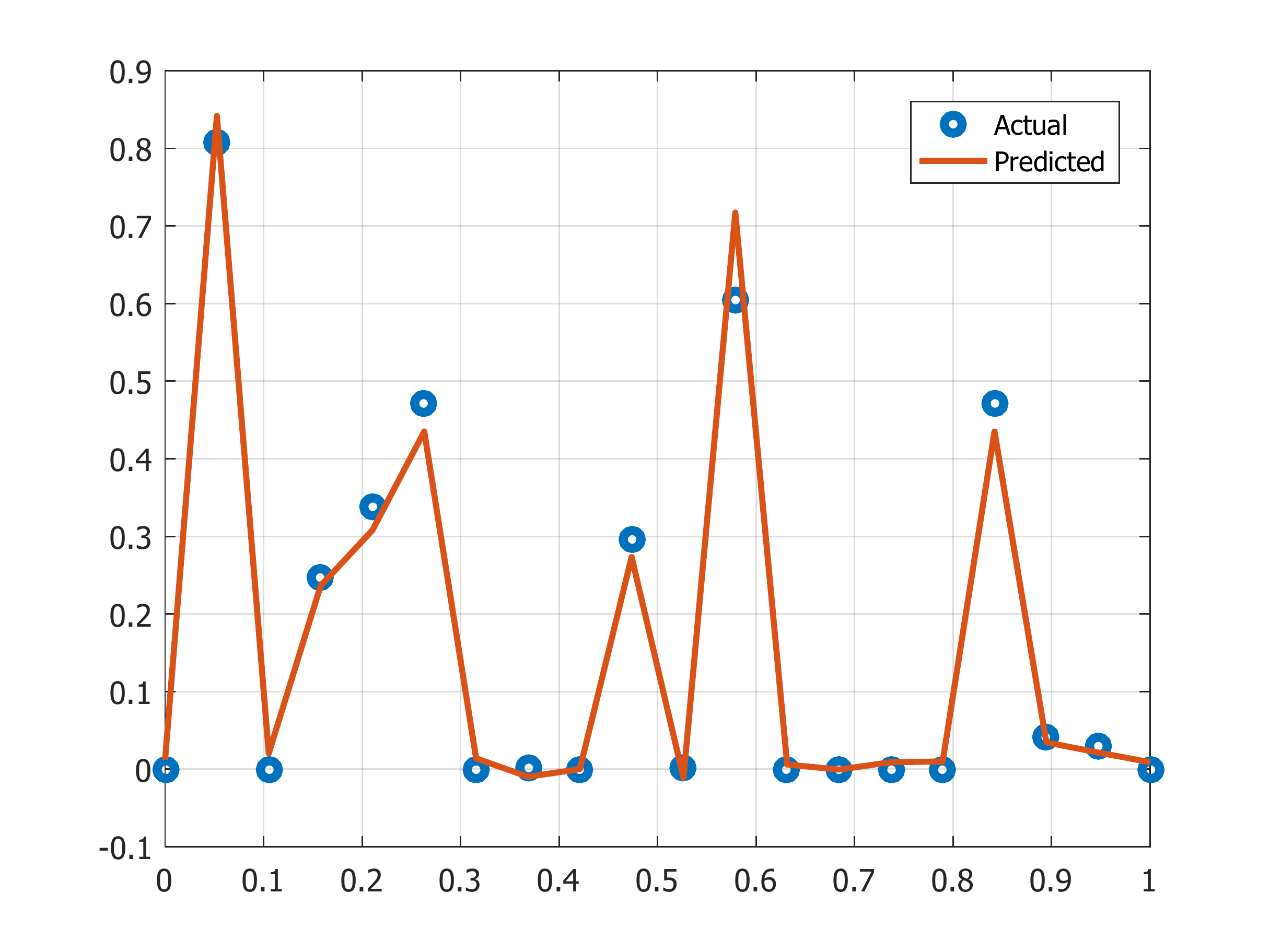} & \includegraphics[scale=0.25]{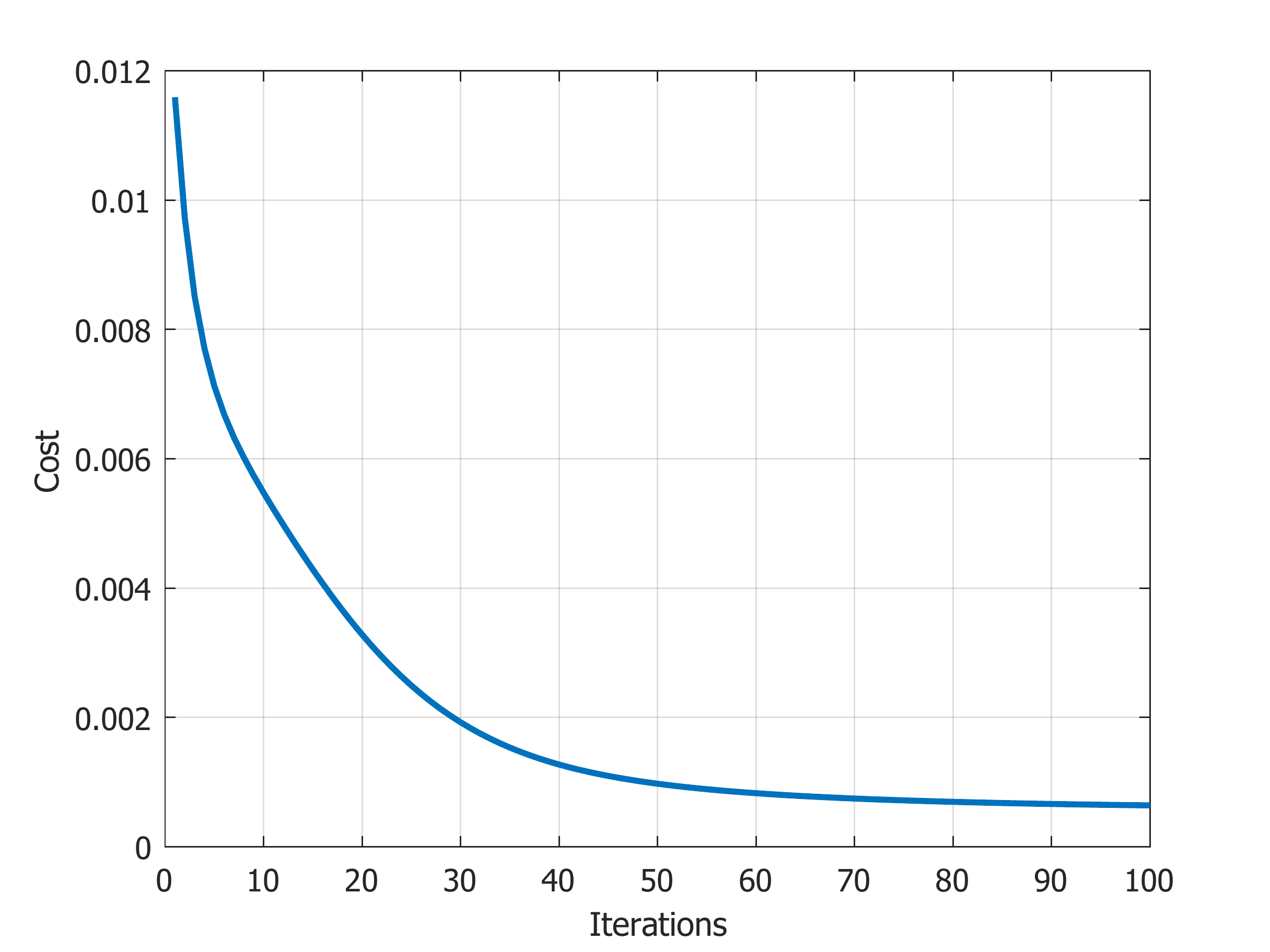}\tabularnewline
			(c) India : Confirmed cases & (g) Cost function decay (c)\tabularnewline
			\includegraphics[scale=0.25]{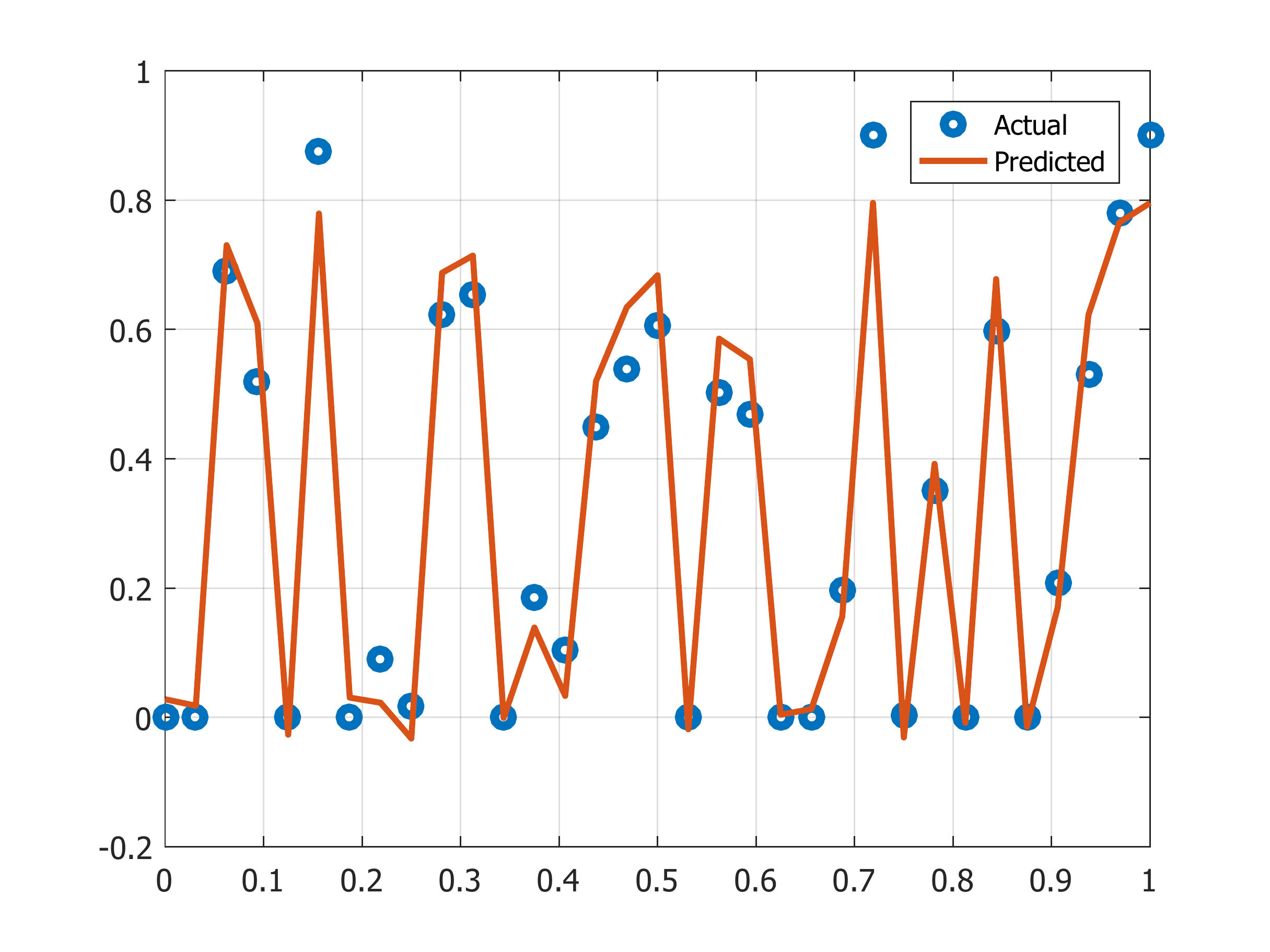} & \includegraphics[scale=0.25]{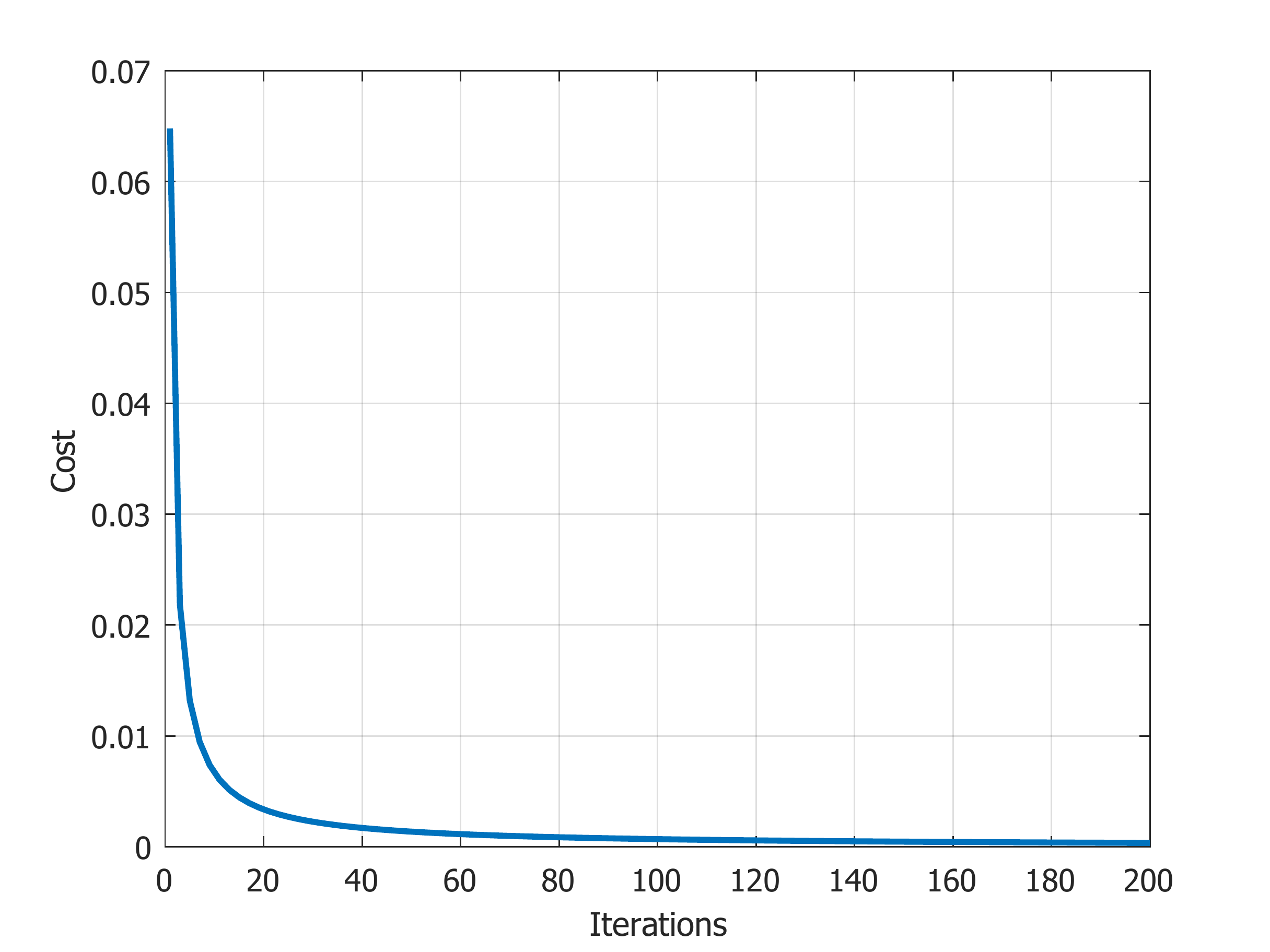}\tabularnewline
			(d) USA : Confirmed cases & (h) Cost function decay (d)\tabularnewline
		\end{tabular}
		\end{centering}
	\caption{Results and cost function decays of CVQNN for number of deaths and
		confirmed cases for COVID-19}
	
\end{figure}

\end{document}